\documentclass[aip,jap,preprint]{revtex4}

\usepackage{graphicx}

\begin{document}
  \title{ Double step structure and meandering due to the many body interaction at GaN(0001) surface in N-rich conditions.}
\author{Magdalena A. Za{\l}uska--Kotur }
\email{zalum@ifpan.edu.pl}
 \affiliation{Institute of Physics, Polish Academy of Sciences,
Al. Lotnik{\'o}w 32/46, 02-668 Warsaw, Poland and Faculty of Mathematics and Natural Sciences,
Card. Stefan Wyszynski University, ul Dewajtis 5, 01-815 Warsaw, Poland}
 \author{Filip Krzy{\.z}ewski} 
\email{fkrzy@ifpan.edu.pl}, 
 \affiliation{Institute of Physics, Polish Academy of Sciences,
Al. Lotnik{\'o}w 32/46, 02-668 Warsaw, Poland} 
\author{Stanis{\l}aw Krukowski} 
\email{stach@unipress.com.pl}
\affiliation{ High Pressure Research Center, Polish Academy of Sciences, ul. Soko\l owska 29/37,01-142 Warsaw, Poland and
Interdisciplinary Centre for Materials Modeling, Warsaw University, Pawi\'nskiego 5a, 02-106 Warsaw, Poland }

\begin{abstract}
Growth of gallium nitride on GaN(0001) surface is modeled by Monte Carlo method. Simulated growth is conducted in N-rich conditions, hence it is controlled by Ga atoms surface diffusion. It is shown that dominating four-body interactions of Ga atoms can cause step flow anisotropy. Kinetic Monte Carlo simulations show that parallel steps with periodic boundary conditions form double terrace structures, whereas initially $V$-shaped parallel step train initially bends and then every second step moves forward, building regular, stationary ordering as observed during MOVPE or HVPE growth of GaN layers. These two phenomena recover surface meandered pair step pattern observed, since 1953, on many semiconductor surfaces, such as SiC, Si or GaN. Change of terrace width or step orientation particle diffusion jump barriers leads either to step meandering or surface roughening. Additionally it is shown that step behavior changes with the Schwoebel barrier height. Furthermore, simulations under conditions corresponding to very  high external particle flux result in triangular  islands  grown at the terraces.  All structures, emerging in the simulations, have their corresponding cases in the experimental results.
  \end{abstract}
%\begin{keyword}
\keywords{
diffusion, lattice gas, surface diffusion , crystal growth}
\pacs{ 02.50.Ga, 81.10.Bk, 66.30.Pa, 68.43.Jk}
%\end{keyword}
%\end{frontmatter}
\maketitle

% main text
\section{Introduction}
\label{sec:A}

Crystal growth remains subject of intensive study, connecting research on fundamental understanding of the  pattern selection, or the kinetic properties and the technical aspects of synthesizing useful materials and structures for present day advanced information technologies, optoelectronics, medicine, or molecular sensing, etc. The shape selection growth phenomena are intimately related to the morphological instability of the flat or curved surfaces of growing crystals \cite{sek1,sek2,langer}. In the macroscale, the variety of the growth shapes extends from flat surfaces, cellular patterns \cite{wol} to dendritic patterns \cite{glik}, and finally to astonishing variety of snow crystal shapes \cite{nakaya,bentley}. The rich variety of the shapes is characterized by different sizes, starting from the above mentioned macroscale patters, to diversity of microcrystals, e.g. spherullitic crystals, and ultimately to nanocrystals, such as fullerenes \cite{ful}, nanotubes \cite{tubes}or graphene \cite{nov}. The shape selection phenomenon encompasses different dimensionality, extending from these three dimensional structures to the various 2-dimensional phenomena, observed on flat crystal surfaces, such as step repulsion, step meandering, and also step bunching. These phenomena, though observed for very long time, still escape from basic understanding. Notably, step meandering, in which the pairs of close steps are bend, interchanging with the other pairs for different direction, observed more than half century ago on silicon carbide surfaces \cite{verma}, was subsequently reported to exist on many other semiconductor surfaces\cite{suna}, such as silicon\cite{verga} or gallium nitride \cite{kruk}. It still remains unsolved puzzle, despite many efforts to understand its dynamics. This work is devoted to elucidation of this phenomenon in the specific case of the growth of gallium nitride by metal organic vapor phase epitaxy (MOVPE) or hydride vapor phase epitaxy (HVPE) on polar GaN(0001) surface \cite{3}.                                                                                               
 
       A model of GaN(0001) surface growing in nitrogen rich condition is proposed. Since, at the standard growth conditions for both MOVPE and HVPE, the system is kept is nitrogen-rich state, Ga atoms motion is considered explicitly as they control the crystal growth process. It is assumed that N atoms are at such abundance at the surface, that they fill lattice sites instantaneously after the incorporation of Ga atoms in the neighboring sites.  The basic assumption, used in the model, is that complete tetrahedral  Ga structure around N atom has lower energy than mere sum of two body Ga--Ga interactions connected via single N atom. It is shown that the presence many-body forces between Ga atoms plays dominant role in the crystal growth dynamics. Strongly connected tetrahedral structures lead to the substantial difference in the speed of the step flow. Growth of wurtzite GaN crystal is modeled by kinetic Monte Carlo (MC) simulations in which stationary step patterns are identified. Using results of the MC simulations we demonstrate that selection of various stationary surface step patterns is realized, depending on the assumed interaction energies and the kinetic growth parameters.

In the proposed model bulk gallium nitride crystal is built by subsequent deposition of consecutive Ga-N atomic layers on GaN(0001) surface, forming stationary step patterns during growth simulations. Initially straight parallel steps, which are oriented along one of the main crystallographic  directions, form double terrace structure when tightly fixed by periodic boundary conditions. In the case when the parallel steps are initially bent  at 120$^o$ forming the $V$-shape form,  they change their orientation to more sharp one first and  then they build double step structure. In the case of initial, larger scale, hexagonal structures, the tendency of step reorientation leads to the characteristic surface wavy pattern (step meandering). In addition to such evolution it is also shown that for high Schwoebel barrier the steps become unstable and such unstable step pattern ultimately evolves to random, rough surface.

\section{The model}
\label{sec:B}
       In the MOVPE and HVPE growth of GaN, the fluxes of the reactive species at GaN(0001) surface, are not equivalent or even comparable. Ammonia flux is typically two (HVPE), or three (MOVPE) orders of magnitude larger than flux of gallium containing molecules. At the typical growth temperature which, for both methods, is close to 1300K, Ga transporting agents, TMG or GaCl, are highly unstable. They decompose  at the GaN surface easily, leaving Ga atoms. Nitrogen transporting agent, ammonia is also unstable. Differently from gallium transporting agents, ammonia decays in the gas phase that creates highly nonreactive molecular nitrogen and hydrogen, reducing availability of the active nitrogen at the surface. Thus larger ammonia flux is needed to flush out these nonreactive decomposition products. 

\begin{figure}
\includegraphics[width=12cm,angle=0]{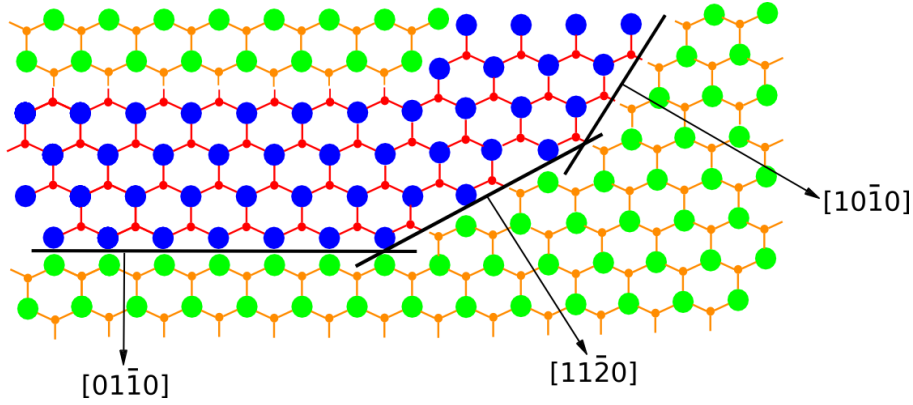}
\caption{\label{model} (color online) Model of GaN crystal. Two types of large circles correspond to Ga atoms at two terrace types, differing by step velocity. From bottom to top terrace heights increase by one layer at each step. Positions of N atoms are marked by dots.}
\end{figure}
\begin{figure}
\includegraphics[width=6cm,angle=-90]{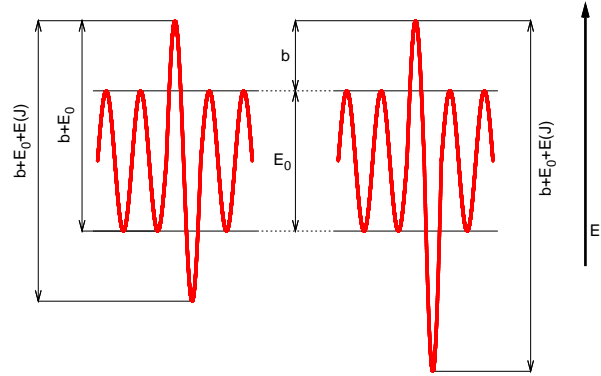}
\caption{\label{potential1} (color online) Potential felt by a particle close to the two step types. Steps go up from right to left side. Particle that attach step from upper step has to overcome Schwoebel barrier. Steps differ by bonding energy, what  decides about the ability of a particle to detach. }
\end{figure}

       It was shown recently that ammonia is attached at GaN (0001) surface in molecular form without energy barrier \cite{13,14,15}. The fast adsorption process of NH$_3$ molecules is accompanied by desorption of hydrogen in molecular form (H$_2$). Thus the entire surface is covered by NH$_2$ radicals, at which, any Ga atom, arriving at the surface, can be attached\cite{16}. Thus the creation of GaN layer proceeds via accumulation of Ga atoms on top of layers of these radicals. 
       
       Due to crystallographic structure of wurtzite, two consecutive dia-atomic layers differ by the location of sites, which are occupied by Ga atoms and by Ga-N bond orientations. Third layer returns to the original position as it is reflected by stacking ABABAB sequence of wurtzite. Due to this the two consecutive surface steps have different atomic structure. As illustrated in Fig \ref{model} three bonds connecting Ga atom with the nearest N atoms within one layer are rotated by 60$^o$ with respect to the bonds in the consecutive Ga layer. As a consequence of this crystallographic buildup, every second step has different atomic structure. Rotation of the step orientation by 60$^o$ exchanges that two kinds of step. Such crystallographic arrangement can result in different step motion under given supersaturation, depending on the step precedence and also on its orientation. 
       
       More detailed analysis of the structure of GaN surface indicates that Ga adatoms, at both types of the step, have the same number of first neighbors (atoms N) and second neighbors (atoms Ga). It follows then that  the step difference cannot be demonstrated within the model, limited to additive two-body interactions only. It turns out that the main difference between both steps lies in the relative orientation of the second neighbors of Ga atom, i.e. the closest Ga adatoms, those from the layer below and those within the layer where Ga atom is located when attached at the surface. The simplest way to model such difference is to introduce four-body interaction between Ga atoms. GaN wurtzite lattice can be build of tetrahedrons of Ga atoms, centered on N atoms. Three atoms of such Ga tetrahedron are shown in Fig \ref{model}, surrounding any displayed N atom there and the fourth Ga atom is located in the layer underneath (not shown there). Alternatively, the N based tetrahedrons with Ga atom inside can be also created, recovering GaN lattice, but this will be not used in the present paper as we are interested in the growth model in which motion of Ga atoms is growth  controlling factor. 
       
       In four-body interaction, each full Ga tetrahedron, bonded by single N atom inside, has different energy than the simple sum of two-body Ga-Ga bonds, connected by N atom (i.e. when the third Ga atom is missing). Note that by assumption of overwhelming presence of nitrogen, we simplified our model, reducing the role of N atoms to the links connecting Ga atoms in the lattice. Also we assumed that the Ga atoms underneath is always present while these in Ga layer above are always empty. Accordingly  our system is described by modeling the energy and dynamics of the single layer Ga atoms only. 
       
       Accordingly to the described buildup of GaN lattice from the Ga tetrahedrons centered on the N atoms, we calculate the given tetrahedron contribution to the energy of Ga surface atom, that depends on the number Ga neighbors, connected by N atom in the tetrahedron center, in the following way:
\begin{equation}
\label{en_od_trojki}
\epsilon_i=\left\{\begin{array}{ll}
1,\quad \textrm{when tetrahedron has all atoms;} \\
\frac{1}{3} r\eta,\quad \textrm{when tetrahedron has empty sites,} \\
\end{array} \right.
\end{equation}
where $\eta$ is a number of Ga occupied neighboring sites, belonging to  a given tetrahedron and $r$ describes the relative  strength of the four-body and the two-body interactions in the system. When $r=1$ two body Ga-Ga interactions sum up to the value characteristic for fully occupied tetrahedron i.e. no additional four-body Ga interactions are present in the system. When $r<1$ three pair bonds, to the nearest neighboring Ga atom of a given Ga in tetrahedron, do not sum up to an energy of the multiparticle configuration. In such a case, the tetrahedron energy is not a simple sum of two body interaction energies only and this is the assumption which is used below. 

At GaN(0001) surface, each Ga surface atom  belongs potentially to four tetrahedrons, three in the  present layer and the one above (practically always empty). Its total energy is
\begin{equation}
\label{en_czastki}
\alpha(J)=J\sum_{i=1}^{4}\epsilon_i.
\end{equation}
where parameter $J$ scales energy of bonds, and the sum runs over four tetrahedrons, surrounding every Ga atom. 
 
It is assumed that, in the surrounding vapor, at the distances comparable to distance between steps, the differences of the concentration of Ga transporting agent are negligible, therefore the Ga atoms are adsorbed at the surface uniformly. The Ga adsorption is accounted for by creation of an adatom at any empty adsorption site, at each MC step, with a probability  
\begin{equation}
\label{p_a}
p_a=\nu_a e^{-\beta\mu},
\end{equation}
 where $\mu$ is a chemical potential, $\nu_a$ is an attempt frequency for the adsorption process and $\beta=1/k_BT$, $T$ is a temperature and $k_B$ is Boltzmann constant. Each adsorbed particle diffuses  over the terrace  until it is attached at the steps. Thus, the possibility of reevaporation is neglected. Probability of a jump from the initial to the final site, in the diffusional movement, is given by
\begin{equation}
\label{p_d}
p_d=\nu_d e^{-\beta (E+E_D)},
\end{equation}
where $E_D$ is energy barrier for diffusion, $E$ depends on the initial $\alpha_i(J)$ and the final $\alpha_f(J)$  energies of the Ga diffusing atom, given by equation (\ref{en_czastki}), in the following manner: 
\begin{equation}
\label{eq:E}
E=\left\{\begin{array}{ll}
\alpha_i(J)-\alpha_f(J) ,\quad \textrm{if $\alpha_i(J)>\alpha_f(J)$;} \\
0, \quad \textrm{otherwise.} \\
\end{array} \right.
\end{equation}
and $\nu_d$ is an attempt frequency for the diffusion process. We use $\nu_d \exp(-\beta E_D)=1$ and $\nu_a=10^{-4}$. The first of these parameters sets the timescale of the whole process. Results of simulations  can be easily rescaled to any other timescale. The second value relates timescale of adsorption process on the timescale of the diffusion process. Note that having $\nu_a$ fixed, we control rate of the adsorption process by setting value of $\mu$. 

Incorporation of Ga adatoms at the step is determined by the rate at which, the Ga adatoms, once adsorbed at the terrace jump to the sites at the step. This rate is additionally modified by Schwoebel barrier \cite{Schwoebel} that sets up different probabilities for atoms jumping to the step from upper and lower terraces. The Schwoebel barrier height is determined by the parameter $B$. Probability of the jump, over the barrier to the step, is given by:
\begin{equation}
\label{p_d^B}
p_d^B=e^{-\beta B}p_d.
\end{equation}
so that the height of the barrier $B$ modifies the standard jump rate, given by equation (\ref{p_d}). The potential felt by a single particle at both step types is sketched in Fig \ref{potential1}. As typically assumed, we impose additional Schwoebel barrier for the jumps from upper terrace. In order to keep the number of control parameters as low as possible we use the same barrier height  $B$ for both types of the steps. In general Schwoebel barrier can depend on the kind of the step, but in order to have similar influence on the step kinetics it should depend on the step orientation also. Technically that means that the barrier should depend on the energy of the particle adsorbed at the step, but such dependence is in our model already accounted for as probability of the jump out the step. We show below that such assumption, being the most natural one, is sufficient to reproduce various variants of the surface kinetics. 

Crystal surface microstate is given by setting the two uppermost layers of atoms. When new step appears, the upper layer is converted into the lower one, and a new layer is built on top of the terrace. The lower terrace is shifted into the bulk of the crystal, not modeled here. In such a way continuity of particle-particle interaction at the step is guaranteed. Every second layer of Ga atoms has different bond orientation. Due to different positions of N atoms, orientation of tetrahedral bonds is rotated by 60$^o$ in alternate layers. Such geometry causes step flow anisotropy characteristic for GaN surface. Step flow anisotropy in the case of Eq. (\ref{p_d}) is realized by the particle jump in the potential that is illustrated in Fig. \ref{potential1}. Note that it differs from the potential proposed in Ref.\onlinecite{Xie}. In our model it is  potential minimum at the step, which is modified by the interactions with other particles, whereas in Ref. \onlinecite{Xie}  the barrier height $B$ for a jump at the step is modified in order to realize different step velocities. 
As a result when $r \neq 1$ we get different bonding not only at every second step for also for the step orientation at the misoriented surface (0001) of GaN crystal.  For the same density of adatoms, the step velocity depends on the particle interaction and therefore it changes with its location and orientation. 

We start our simulations with an  even  number of equally spaced, parallel steps  on the surface. Heights of the neighboring steps differ by one Ga atomic layer. New Ga atoms are adsorbed at the surface. Periodic boundary conditions are applied in the lateral direction while in the direction in which the crystal grows they are corrected by constant height difference between both sides of the simulated area.

\begin{figure}
\includegraphics[width=12cm,angle=-90]{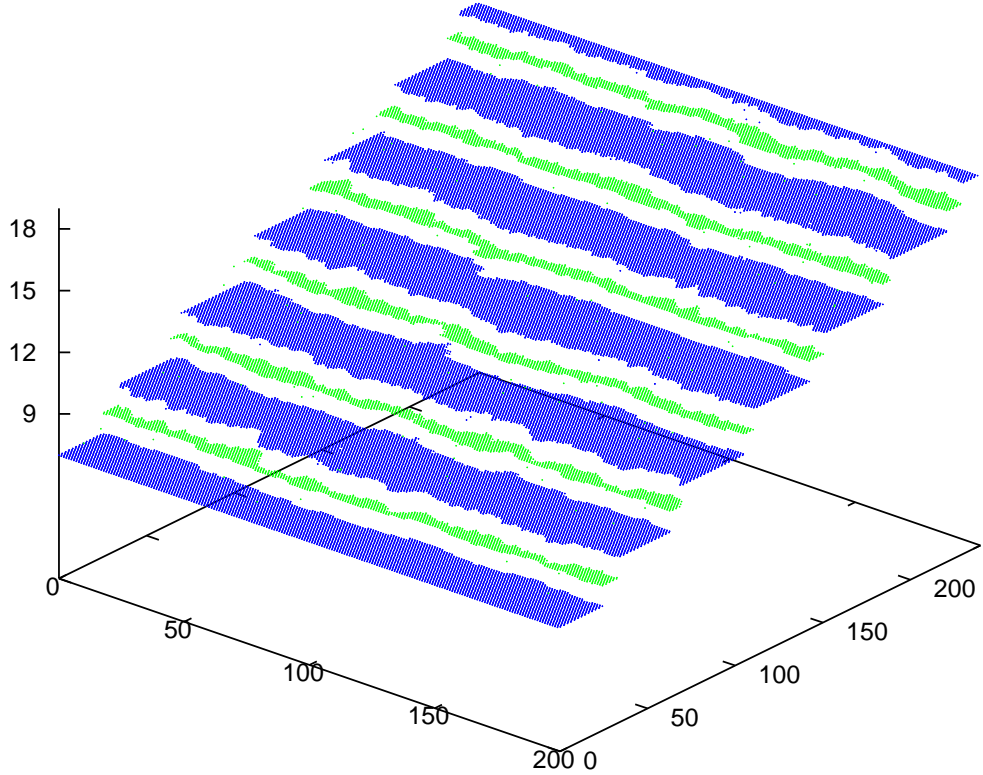}
\caption{\label{3d} (color online) Stationary pattern of double  terrace structure of different width and straight steps. The initial terrace width was $20$ lattice constants. Simulation was carried out for system of size $200 \times 240$ lattice constants, $r=0.4$ , $B=0$, $\beta J=5$ and $\mu=15$. Steps are perpendicular to $[01\bar{1}0]$ orientation.
}
\end{figure}

\begin{figure}
\includegraphics[width=12cm,angle=0]{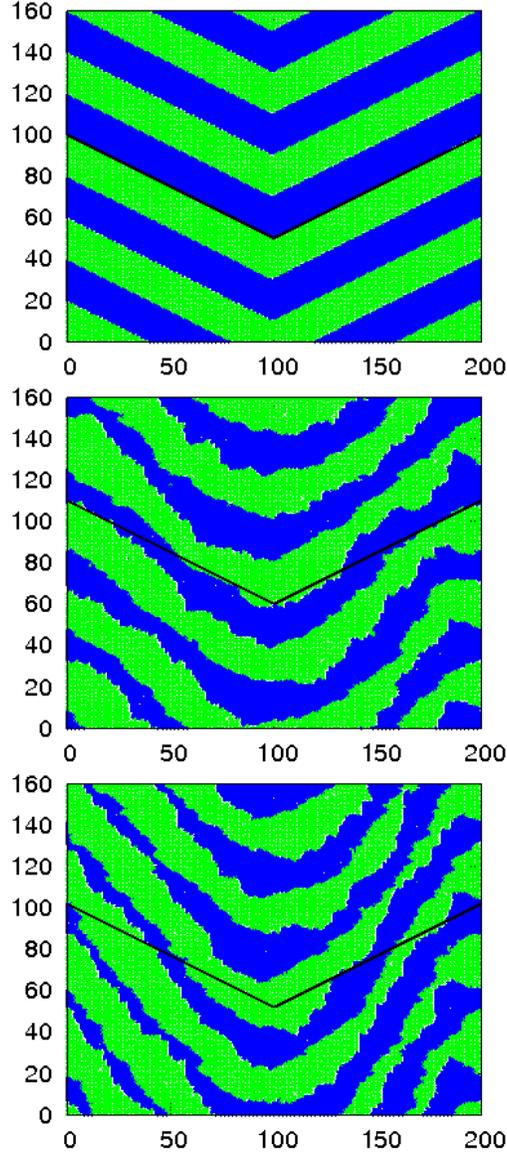}
\caption{\label{regular} (color online) Step pattern evolution for system with steps initially oriented perpendicularly to $[10\bar{1}0]$ and $[01\bar{1}0]$. Surface patterns of a system for following  evolution times $0$, $1^.  10^7$, and $3^. 10^7$ MC steps
are presented from top to bottom. Step anisotropy is given by $r=0.4$,  $\beta \mu=15$, $\beta J=5 $, $B=0$ and at it the system of size $200 \times 160$ lattice constants,  eight $V$ shaped terraces were prepared. For comparison, initial step shape is drawn in thin black line.}
 \end{figure}

\section{ Step doubling }
\label{sec:B2}

\begin{figure}
\includegraphics[width=12cm,angle=0]{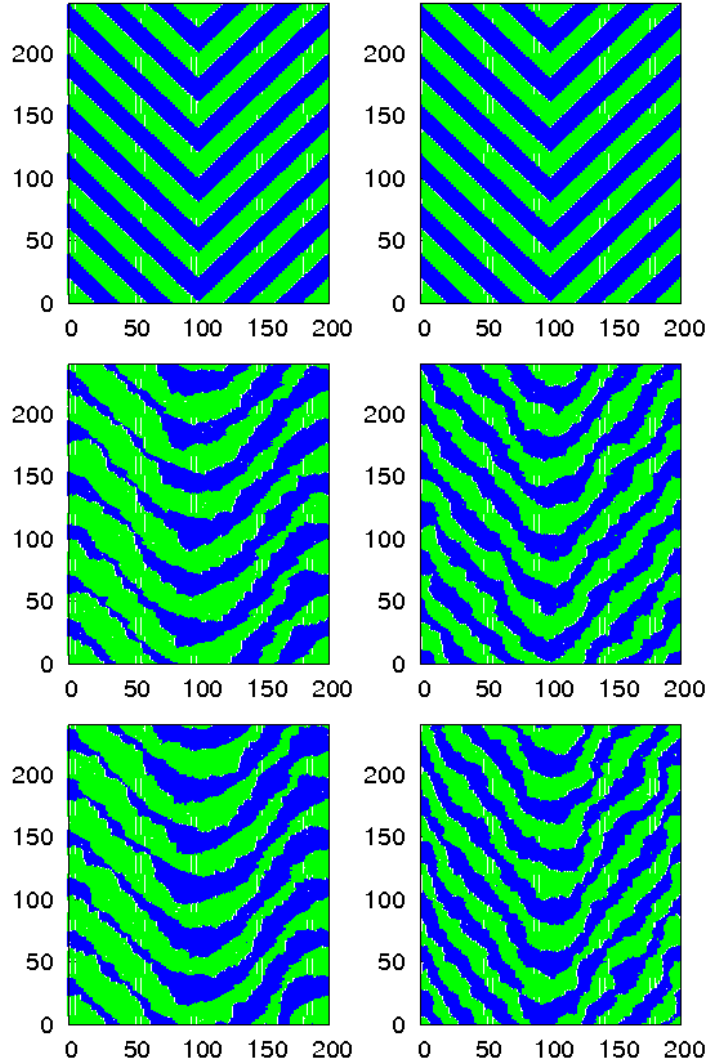}
\caption{\label{stat} (color online) Step pattern evolution for systems with  anisotropic and isotropic steps. Initially steps were bend $19^o$ to the crystallographic directions. Consecutive evolution plots are presented from top to bottom for the system with anisotropic steps described by parameter $r=0.4$ at left hand side and isotropic steps, $r=1$ at the right hand side. System was $200 \times 240$ lattice constants with 12 $V$ shaped terraces and other parameters: $\beta \mu=15$, $\beta J=5 $, $B=0$ Number of MC steps changes from top down as follows: $0$, $5^.  10^6$, and $10^7$}
\end{figure}

We begin our studies from  the case starting from the initial configuration, having straight terraces of equal widths  and the steps descending  along $[01\bar{1}0]$ direction, i.e. the steps are oriented in $[11{\bar2}0]$ direction. After very short initial time, around $10^5$ MC steps, the step system evolves to the stationary double terrace pattern, shown in Fig. \ref{3d}. Faster steps move forward, and in the result, every second terrace shrinks, being close to disappearance. Such pattern is very stable and it is preserved, essentially in the same form, during simulations building up to eight consecutive layers, as illustrated in Fig \ref{3d}. The height of the lowest terrace of the modeled area was initially set to zero, thus its displayed height indicates the number of newly deposited layers during simulation. Let us stress that the stationary state like the presented one was attained only for the step anisotropy caused by four-body Ga interactions. For the isotropic case (r = 1), the width of the terraces does not change, indicating the same kinetics of both steps. The parameter range for which  system evolves into such pattern is determined by the complex interplay of the temperature, particle flow, Schwoebel barrier height and the anisotropy. In the case when the terraces are wider or for higher particle flux, for given temperature and anisotropy, the new phenomenon of the step meandering emerges. 

As proposed initially by Burton, Carbera and Frank (BCF), dislocations of the screw or mixed type, serve as continuous step sources on otherwise atomically flat crystal surfaces \cite{BCF}. Thus it could be assumed that steps are created at the location where such dislocation line emerges at the surface. Under supersaturation, the steps attain the spiral shapes, extending over large part of the surface. The exact spiral shape depends on the anisotropy and the temperature, being rounded, close to Archimedean, for high temperatures, and having straight line fragments for low temperatures. GaN growth corresponds to low temperature case in such fashion that the steps, limiting flat terraces, encircle the dislocation line, creating hexagonal pyramid. Thus these steps are directed along  six sides of hexagon. 

In order to model such case we prepared array of V-shaped steps, broken in the center, with its parts oriented along two neighboring sides of the hexagon, i.e. along the following directions: $[10{\bar1}0]$ and $[01{\bar1}0]$. Such design could be treated as a segment of a step  pyramid, corresponding to two sides of the hexagon. Such initial configuration of the system is depicted in top panel in Fig.\ref{regular}. Both halves of this system create regular pattern of parallel steps, like initial configuration in the previous case. However, the plots showing time evolution of the step pattern, shown in  Fig.\ref{regular}, prove that the system behavior is completely  different from that presented in Fig. \ref{3d}. In place of the double parallel  straight step terrace structure, the steps have tendency to bend in such a way that $V$-shape  becomes more sharp. Steps move closer, but their motion is not the uniform, some moves faster, the other slower, in the result they start to form meanders. Slowly they bend to different orientation, evolving into one which seems to be the stationary state for the system. When proper orientation is reached the step system evolves to the double terrace structure, preserving the selected step orientation, at the approximate angle $19^o$ to main, initially used  crystallographic direction. It is worth mentioning that the terraces widths are not the same, green and black terraces being wider on the left and right side, respectively.

In order to find out whether such curved step line is dynamically stable, $V$-shaped configuration, we started our simulations with the initial pattern of steps, oriented at angle $19^o$ to the main crystallographic direction, as  shown in  Fig.\ref{stat}. With steps initially bent the system  evolved  into stationary state presented at the left side of the Fig.\ref{stat}. System with anisotropic steps forms patterns  of wide and narrow terraces following each other. When the step changes its orientation, wide terraces smoothly evolve into narrow ones building characteristic  pattern, which is often observed experimentally at the surface of MOVPE grown GaN layers \cite{kruk}. The step pattern, obtained in the case for which the steps are  anisotropic, plotted at left hand side of Fig. \ref{stat}, could be compared with the one obtained for the isotropic case, plotted at the right hand side of Fig. \ref{stat}. In the latter case, the terraces are equally spaced and therefore, for isotropic case, dynamically stable step pattern is completely different. 

It is interesting to find out which factor is responsible for such difference in the step evolution from a set of parallel steps, shown Fig \ref{3d} and from $V$-shaped pattern, demonstrated in Fig \ref{regular}. It seems that the difference between these two cases stems from initial conditions coupling to boundary conditions. In the first case, the steps are fixed by periodic boundary conditions, making the whole structure very rigid. Each step, attached at both sides, could move forward only, and finally characteristic periodic step pair structure is formed. It is possible that for larger size, the system will evolve into the $V$ shaped pattern, observed for the second case, where periodic boundary conditions are also applied, but $V$ shaped structure leaves freedom for step bending, and it appears that the steps change their orientation to the one with more sharp apexes. In the context of hexagonal patterns, formed around the dislocation line at the beginning, when steps are short, they are rather in rigid state, each kept by two apexes formed in such a way that they cannot sharpen at the same time. In such situation every second step moves forward creating double step structure. However when the structure grows further and step extends, the pattern is no longer so stiff and starts to bend, forming characteristic wave-like structures. Thus both types of behavior together can explain example of characteristic rosette shapes, observed around spiral dislocation \cite{kruk}.

\begin{figure}
\includegraphics[width=6cm,angle=-90]{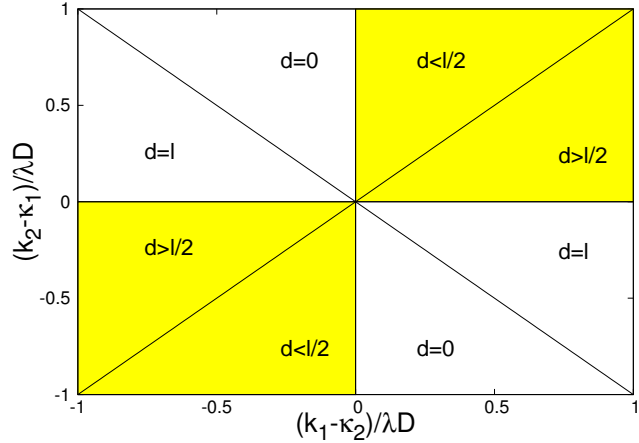}
\caption{\label{phase} (color online) Phase diagram for step patterns as given by Esq. (12)and (13) plotted in $k_2-\kappa_1$ vs. $k_1-\kappa_1$ variables. For parameters which lie within painted regions of the plot the relative width of terraces $d/l$ is given by the solution of Eq. (12). For other parts of the parameter space, one of the terraces disappears. }
\end{figure}

\section{ BCF analysis of anisotropic steps }
\label{sec:C1}
The mechanism of a double step pattern creation, in the system having two types of anisotropic steps can be analyzed employing Burton-Carbera-Franck (BCF) model \cite{BCF} of the surface diffusion controlled growth. In the BCF model it is assumed that due to the incorporation of the adsorbed molecules at the steps, its local density $\rho$ changes, being the smallest at the steps. Accordingly, the particles diffuse towards the steps. In the simplest formulation one can consider a train of parallel steps, i.e. the problem is reduced to a one dimensional diffusion system. In the case of a stationary step motion with constant velocity $V$, we the mass balance equation, in the step coordinate system, can be written in the time independent form:
\begin{equation}
\label{BCF}
D \frac{d^2}{(d z)^2} \rho + F - \frac{\rho}{\tau}+ V \frac{d}{d z}\rho=0
\end{equation}  
where $D$ is the diffusion  coefficient, $F$ describes the flux of incoming atoms, adsorbed at the terrace and $\tau$ is an average lifetime of an atom in a state of mobile adsorption on the crystal surface. In the MC model presented above, adatoms do not desorb, but double occupancy is forbidden that leads to the relation $\tau=1/F$.
 Equation (\ref{BCF}) is formulated in the frame attached to the step that moves with constant velocity $V$, hence the latter, velocity dependent term arises from Galilean transformation from crystal lattice to step attached coordinate system. We express local density of the adsorbed particles $\rho$ in units of a number of particles per one lattice site, and similarly all distances are measured in a number of intersite distances (i.e. lattice constants $a$). Thus $z$ is dimensionless and the unit of $D$, $F$ and $V$ is $1/s$. Solution  of this equation (\ref{BCF}), has the form 
\begin{eqnarray}
\label{solution}
\rho(z)& = & F \tau - A \cosh(\lambda z) - B\sinh(\lambda z)  
\end{eqnarray}
where, for small $V$:
\begin{equation}
\label{lambda}
\lambda = \sqrt{\frac{1}{D \tau}}  
\end{equation}
The choice of the boundary conditions completely specifies problem. With two different step types we 
have to assume two density profiles (\ref{solution}) with different constants $A$, $B$. For each of density profiles we have two edges, what gives four boundary conditions. Let the length of the first of  two consecutive terraces be equal to  $d$, and of the second to  $l-d$. Thus we assume that the double step structure repeats with period length $l$ , i.e. the surface is covered by an infinite train of parallel steps, at which the following boundary conditions are imposed on the density of adatoms at the terraces: 
\begin{eqnarray}
\label{boundary}
 D \frac{d\rho}{d z} \mid_{(-d)+} & = & k_1(\rho-{\rho^+}_1) \mid_{(-d)+} \nonumber  \\
 -D \frac{d\rho}{d z} \mid_{0-} & = & \kappa_2(\rho-{\rho^-}_2)\mid_{0-} \nonumber \\
 D \frac{d\rho}{d z} \mid_{0+} & = & k_2(\rho-{\rho^+}_2) \mid_{0+}  \\
 -D \frac{d\rho}{d z} \mid_{(l-d)-}& = & \kappa_1(\rho-{\rho^-}_1) \mid_{(l-d)-} \ \nonumber
\end{eqnarray}

In order to make our calculations simpler we position one step at $z=0$, and the other one at $-d$, which is periodically equivalent to $l-d$. Parameters $k_1, \kappa_1$ and $k_2, \kappa_2$ describe the rates with which particles attach from the right and left side to the  first and to the second step and ${\rho^{\pm}}_i$ is equilibrium density at step $i$.  The velocity of the step motion should be equal for both steps, and is given by
\begin{equation}
\label{V}
V =  D \frac{d\rho}{d z} \mid_{0+}+D \frac{d\rho}{d z} \mid_{0-}= D \frac{d\rho}{d z} \mid_{(-d)+} +D \frac{d \rho}{d z} \mid_{(l-d)-}  
\end{equation}
When conditions (\ref{boundary}) and (\ref{V}) are imposed, the solution (\ref{solution}) it turns out that
the width $d$  can be determined from the relation
\begin{eqnarray}
\label{a}
\frac{k_1+\kappa_2}{k_1-\kappa_2}\coth(\lambda d)- \frac{k_2+\kappa_1}{k_2-\kappa_1}\coth[\lambda(l-d)]= \\
\nonumber \frac{\kappa_1 k_2+\lambda^2 D^2}{\lambda D (k_2-\kappa_1)}- \frac{\kappa_2 k_1+\lambda^2 D^2}{\lambda D (k_1-\kappa_2)}
\end{eqnarray}
 
\begin{figure}
\includegraphics[width=12cm,angle=0]{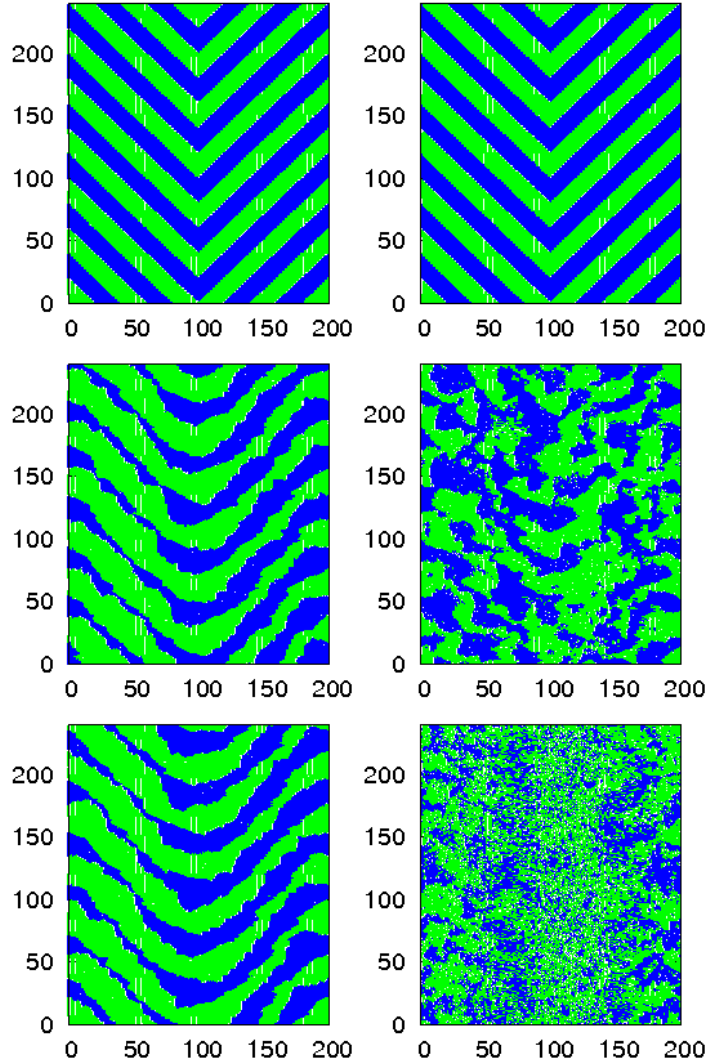}
\caption{\label{bar} (color online) Step pattern evolution for two different Schwoebel barriers. Steps initially were oriented as in Fig.\ref{stat}. Consecutive evolution plots are presented from top to bottom for the system with low  Schwoebel barrier $\beta B=0.5$ at left hand side and with $\beta B=1$ at right hand side, other parameters of the systems were   $\beta \mu=15$, $\beta J=5 $,initial step width equal to 20 lattice constants and number of MC steps changes from top down as follows: $0$, $5^.  10^6$, and $10^7$}
\end{figure}

The above equations describe complete solution of the stationary dynamics of the double step structure, obtained from Eq.(\ref{BCF}). Step permeability was neglected here. It general, it can be accounted for as in Ref.\onlinecite{rang},  slightly complicating  the above expressions but not changing the conclusions.  The solution of Eq. (\ref{a}) can be obtained providing that the  condition
\begin{equation}
\label{cond}
(k_1- \kappa_2)(k_2-\kappa_1)>0 
\end{equation}
is fulfilled. In such case Eq.(\ref{a}) determines the relative widths $d$ and $l-d$ of two consecutive terraces in a stationary solution.   
However when $\kappa_2$ becomes close to $k_1$ or $\kappa _1 \approx k_2$ the first or the second type of terrace disappears, the condition (\ref{cond}) is not obeyed, and Eq.(\ref{a}) does not describe relative terrace widths. Double steps form and they propagate differently from a pairs of steps described above. 

In Fig. \ref{phase} we plot regions of parameters for which different step behavior is observed. Double step structure is observed when the terrace width $d=0$ or $d=l$. This was the case, described in the previous Section. When the Schwoebel barrier is zero i.e. when the rates of the particle attachment to the step from upper and lower terrace are the same, $k_1=\kappa_1$ and $k_2=\kappa_2$, the condition (\ref{cond}) is not fulfilled. In such case, the faster step overtakes the slower one, and double step structure forms. In the special case, when all rates are the same, the steps move with the same speed, thus the terrace widths do not change. They stay essentially identical to the initial state. Note that in simulations presented in Figs. \ref{3d}, \ref{regular} and \ref{stat}, every second  terrace is very  narrow, but it does not disappear totally. This can be due to the discrete character of the system, interactions between diffusing particles and the fact that the system is not one-dimensional, but has the second dimension parallel to the step. The presence of the second dimension can  play an important role in the system kinetics. Condition (\ref{cond}) is fulfilled for systems with high Schwoebel barrier for which relation (\ref{a}) describes relative terrace widths in a pair step pattern.    

When Schwoebel barrier is very high, $\kappa_1=\kappa_2=0$ equation (\ref{a}) simplifies to the formula \cite{JNS}
\begin{equation}
\label{abb}
\coth(\lambda d)-\coth[\lambda(l-d)]= D \lambda (\frac{1}{k_2}-\frac{1}{k_1})
\end{equation}
This equation could be solved for any value of the pair of parameters $k_1, k_2$, hence this system always attains stationary state, with two terraces having different, finite widths.

\begin{figure}
\includegraphics[width=12cm,angle=0]{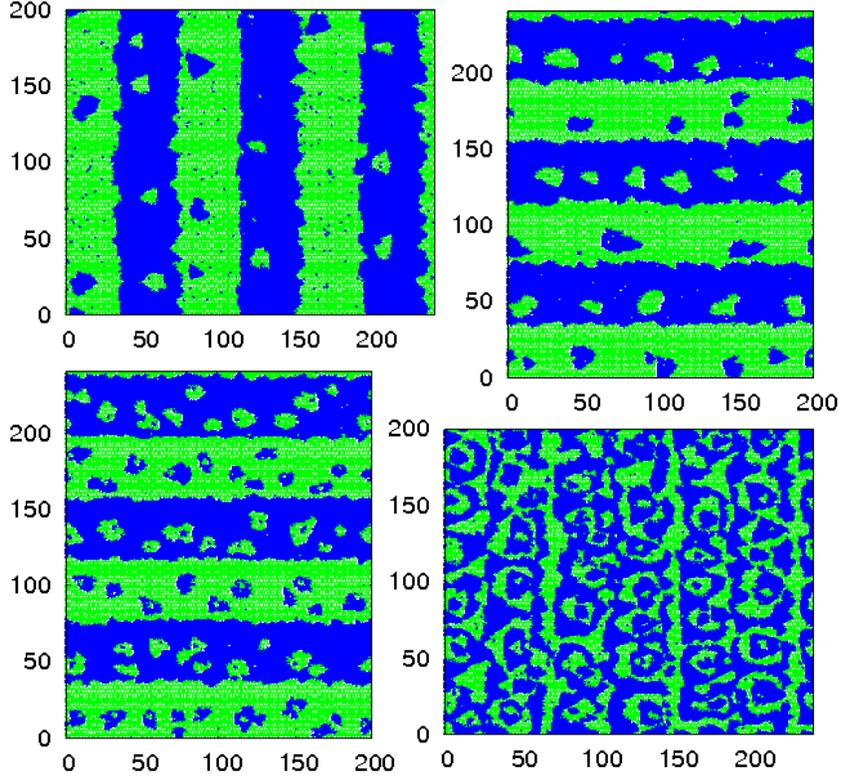}
\caption{\label{domains} (color online) Domains  for large particle flux, $\mu=13$. Upper row shows system with $r=0.4$ for steps perpendicular to direction $[11\bar{2}0]$  at right hand side and  with steps perpendicular to direction $[10\bar{1}0]$ at left hand side. Bottom row shows isotropic system i.e. for $r=1$ at right hand side and anisotropic system after longer evolution. Three first cases are shown after $10^4$ MC steps and the last one after $10^5$ MC steps. In these simulations $\beta J =5$ and $\beta B=0$ (no Schwoebel barrier).}
\end{figure}

\section{Step meandering }
Step anisotropy in  the system leads to the stable  double terrace structure for sufficiently high disorientation, low particle flow and low Schwoebel barrier.  For wider terraces the steps start to meander and cannot overtake each other. We do not observe double step structure for wider terraces. Instead, patterns of meandering steps change during surface evolution.  The value of terrace width at which meandering starts, depends on the Schwoebel barrier and on the flux of incoming particles. In Fig. \ref{bar} step evolution for the two systems, having different Schwoebel barrier only, is shown. Plotted, at the left hand side of Fig. \ref{bar}, the evolution of the pattern, obtained for the barrier $B=0.5 J$, does not differ much from that for $B=0$, shown in Fig.\ref{stat}. When  $r=0.4$, the barrier $B=0.5 J$  corresponds also to the case of the condition (\ref{cond}) not fulfilled, i.e. the same as for $B=0$ at any flux. For the  barrier increased to above $0.6J$, the growth, in accordance to the fulfilled condition (\ref{cond}), tends to attain stable pattern of the two terraces having different widths. At very high barrier, i.e. increased up to $B=J$, the  steps are destabilized by fragmentation so that the system eventually evolves into the rough surface, the mechanism known as Ehrlich-Schwoebel effect \cite{Schwoebel,rusanen,rusanen2}.  In Fig. \ref{bar}, the patterns obtained for two different Schwoebel barriers at the same evolution times, are compared. Destabilization, due to the particle flux increase, looks similarly to this one. When particles attach step so frequently that the diffusion along step is too slow for step smoothening, meandering is overcome by the step roughening and fragmentation. Note, that beside difference in the step detachment rates we assume no diffusion, or potential anisotropy, which were present in other MC models \cite{kato,kato2}.
 
In the case when the chemical potential of the vapor phase is set very high, i.e. when the external particle flux is very intense, the particles tend to form islands at the terraces (nucleation). The phenomenon occurs because incoming particles reach surface so often that they rather find others, stick  together and create islands (nuclei of new layer), before they could reach steps. In Fig. \ref{domains} in two top plots, we see triangular domains oriented antiparallely at every second terraces, obtained for the two different, perpendicular orientations of the steps. Left hand side top panel shows steps perpendicular to  $[11\bar{2}0]$ direction and right hand side top panel steps perpendicular to  $[10\bar{1}0]$ direction. Both were grown for $r=0.4$ parameter. Left hand side bottom panel shows domains for  the isotropic case, i.e. for $r=1$ at the same growth conditions as in the upper, right diagram. In this case, the nuclei of the new layer are more regular, close to hexagonal. The latter picture, shown in the right bottom, presents rough surface, obtained during longer evolution of an anisotropic system shown the left, upper diagram. 
From these diagram we conclude that the particle density at the terrace, or equivalently, the flux of incoming atoms, should be low enough (low supersaturation) in order to attain stationary step flow.

\section{Stability analysis}
In order to check when parallel steps start to meander we carry out stability analysis \cite{bales,bena,misbah}. In such analysis we perturb the shape of each step with a finite harmonic oscillations of wavevector $k$ and frequency $\omega$. Position of $i$-th step ($i=1,2$) becomes 
\begin{equation}
z_i(x)= {z_i}^0+\delta_i \sin(k x+\omega t)
\end{equation}
where $\delta_i$  is the amplitude of small perturbation of the step position.
The particle density at each terrace up to the first order is given by
\begin{eqnarray}
\rho(x,z,t)&=&A_0 \sinh(\lambda z(x,t)) +B_0\sinh(\lambda z(x,t)) \\
	&+&  \epsilon [A_1\sinh(\Lambda z)	+ B_1 \sinh(\Lambda z)] \sin(k x+ \omega t) \nonumber 	
\end{eqnarray}
with small parameter $\epsilon$  and $\Lambda=\sqrt{k^2+\lambda^2}$ 
In the boundary conditions  (\ref{boundary})  full expression for the equilibrium density should be used\cite{bales,bena}
\begin{equation}
{\rho^\pm}_i={\rho^\pm}_{i0}- \Gamma \frac{{z}_{xx}}{(\sqrt{1+{z_x}^2})^{3}}
\end{equation}
where $\Gamma$ is capillary length of the step \cite{langer}.
The velocity of bent step should be corrected as follows\cite{bales,bena}
\begin{equation}
V=\frac{V_0+\dot{z}}{\sqrt{1+{z_x}^2}}
\end{equation}
We look for such value of $k$ which 
fulfills the following two conditions  \cite{bales,bena}
\begin{eqnarray}
\label{kc}
\omega(k_c)=0 
 \ \ \
\frac{\partial \omega}{\partial k}_{k=k_c}=0
\end{eqnarray}
The conditions determine such relation between parameters that separates stable straight moving steps from that for which unstable modes emerge. 

Let us first analyze the case where $k_1 \approx \kappa_1$ and $k_2 \approx \kappa_2$, like the one presented in Figs \ref{3d},\ref{regular}, \ref{stat},  and in Fig. \ref{bar} at left hand side. These step attachment rates obviously do not fulfill condition (\ref{cond}), so double step structure is formed. Such double step pattern is a uniform system with terraces of width $l$ and the steps having growth rate equal to the slower of $k_1, k_2$. In such a case the solution of (\ref{kc}) in the uniform step system is $k_c=0$, and step instability is found when the following relation between the system parameters is fulfilled 
   \begin{eqnarray}
\label{meandr0}
\frac{\lambda \Gamma}{ F \tau }\big[\coth(\lambda l)\frac{k+\kappa}{k-\kappa} +\frac{(D \lambda)^2+k \kappa}{D \lambda (k-\kappa)}\big]<\frac{1}{2}, 
\end{eqnarray} 
where we assumed that $k=k_1<k_2$ and $\kappa=\kappa_1$. According to the Eq. (\ref{meandr0}) the steps are not destabilized when $\kappa \rightarrow k$. In the classical relations used in Refs \onlinecite{bales,bena}, the value of $\lambda l$ is  assumed to be very large and $k$ infinite. In the limit of large $\lambda l$ and $k$, with $\kappa=0$, a standard condition $ (\lambda \Gamma)/ (F \tau)<1/2 $ defined in \cite{bales,bena}, is recovered. In general, at the  higher flux value $F$, the steps are more unstable. It follows also from a full version of the relation (\ref{meandr0}) that large terrace width $l$ or high step adsorption rate $k$ also lead to step meandering. According to Eq. (\ref{meandr0}), large diffusion coefficient has no influence on step stability, however when  diffusion is very low, its increase can destabilize steps.  

In the case when the condition (\ref{cond}) is fulfilled, the terraces of two different widths exist, terminated by steps having different rates. We assume that at both steps parameters $\delta $ and $\epsilon$ are different. Two different steps are assumed and then two different values of $\delta= \delta_1,\delta_2$ and $\epsilon=\epsilon_1,\epsilon_2$ are present. Moreover as  in  Eqs. (\ref{boundary}) ${z_1}^0=-d, {z_2}^0=0$. Finally the following relation is obtained
\begin{eqnarray}
\label{meandr}
\frac{\lambda \Gamma}{ F \tau }\big[\coth(\lambda d)\frac{k_1+\kappa_2}{k_1-\kappa_2} +\frac{(D \lambda)^2+k_1 \kappa_2}{D \lambda (k_1-\kappa_2)}\big]<\frac{1}{2}, 
\end{eqnarray}

According to Eq. (\ref{a}) the above condition can be formulated for the second terrace width $l-d$ replacing $d$ and for $k_2, \kappa_1$ substituting $k_1, \kappa_2$. When equilibrium conditions (\ref{a}) are fulfilled on the base of (\ref{meandr}) both step destabilize at once.  In fact such behavior is visible in Fig. \ref{bar}.

\section{Conclusions}
\label{sec:D}

At GaN(0001) misoriented surface two crystallographically different kinds of steps are present. The difference stems for the way how particle are bonded at the step. Ga adatoms attach step via one N atom, closing tetrahedral Ga  structure or via two independent N atoms leaving two open structures. We assumed that closed tetrahedral Ga structure has much lower energy than sum of simple Ga-N-Ga bond.   As an effect particle step detachment rates change at every second steps. Rotation of the step  orientation by 60$^o$ exchanges that two kinds of steps. We show that assumption of strong bonding of closed tetrahedral structures has its consequences in the specific stationary surface patters. We have demonstrated that equally distanced parallel steps evolve towards structure of pairs of terraces of different width. For some parameters values, one of terraces reduces to zero and double step structure is build. The condition for such behavior was calculated using a one-dimensional BCF formulation.  In the other considered case, the steps, which initially form $V$-shaped structure change their orientation first, after which every second step moves forward forming pairs of step structure. When $V$-shaped steps are initially prepared along lines of higher crystallographic numbers, they strictly evolve into the double step structures.   Both these types of system behavior together explain step patterns observed  in the growth on many semiconductor surfaces, such as SiC\cite{verma}, Si\cite{suna,verga} and GaN\cite{kruk}. For small structures, that are fixed at both ends, like in spiral structure two opposite step bending, we observe pair of terraces terminated by parallel steps. For longer hexagon edges, the steps bend and we observe wave-like step pair structures. When external particle flow increases or for higher Schwoebel barrier, the step starts meandering and after that surface becomes rough. Stability analysis of double step structure allows to find conditions for which step destabilizes. It appears that both steps destabilize for the same conditions. Therefore the main distinctive feature of the proposed model i.e. the existence of many-body interactions between Ga atoms allowed us to recover various surface GaN(0001) phenomena, observed experimentally \cite{kruk}.  

\begin{acknowledgements}
 Research supported by the European Union within European Regional Development Fund, through grant Innovative Economy (POIG.01.01.02-00-008/08)
\end{acknowledgements}

\end{document}